\documentstyle[12pt]{article}
\begin{document}

\vspace{1cm}
\title{Mean-field calculations of quasi-elastic \\
       responses in $^4$He.}

\author{ M.Anguiano, A.M.Lallena \\
{\small Departamento de F\'{i}sica
Moderna, Universidad de Granada,} \\
     {\small \sl E-18071 Granada, Spain} \\ \\
     and \\ \\
     G.Co'\\
{\small Dipartimento di Fisica, Universit\`a di Lecce and} \\
{\small INFN,Sezione di Lecce,}\\
{\small \sl I-73100 Lecce, Italy }  } 

\date{\mbox{ }}
\maketitle

\begin{abstract}
We present calculations of the quasi--elastic responses
functions in $^4$He based upon a mean--field model used to
perform analogous calculations in heavier nuclei. The meson
exchange current contribution is small if compared with the
results of calculations where short--range correlations are
explicitly considered. It is argued that the presence of these
correlations in the description of the nuclear wave functions 
is crucial to make meson exchange current effects appreciable.
\end{abstract}

\vskip 0.6 cm

The evaluation of meson exchange current (MEC) effects in nuclei
is a topic which has been investigated for more than twenty
years. Various methods have been used to calculate these
effects, and a great variety of nuclei and observables have been
investigated.

A clear fact arising from the large amount of results produced
in these years is that the effects of MEC are large for few body
systems \cite{fro91}, whereas they appear to be rather small in
medium and heavy nuclei \cite{lowen}--\cite{ama94}. 

In a previous work \cite{ama94} we have argued that this can be
ascribed to the presence of short--range correlation functions
in the models describing the few body systems. 

In the case of the deuteron short--range 
correlations in
both non relativistic \cite{car91} and
relativistic \cite{gro91} calculations are explicitly included.
The systems with 3 or 4 nucleons have been studied using different
techniques (Faddeev equations \cite{fri91}, hyperspherical functions
\cite{ros90}, Green function and variational Monte Carlo
\cite{car91}, etc.) but all of them consider these correlations. In
these few body systems, the MEC produce large effects at any energy
scale considered, either in the ground state observables
\cite{fro91} or in the quasi-elastic  response \cite{car94} and even
at higher energies.

For medium--heavy nuclei, nuclear models which take into account
short--range correlations have been recently proposed
\cite{pie90}. The present status of the
art in this field is however quite far from the possibility of
calculating MEC contributions. 
The effects of the MEC in these nuclei, either in the ground
and low-lying states \cite{lowen}--\cite{ama92} or in the case of
nuclear excitations in the continuum 
\cite{mediu}--\cite{ama94}, have been evaluated  within the
mean--field approach. 
Contrary to what has been found in the few body systems, 
in medium--heavy nuclei these effects
are rather small, i. e. they are of the
same order of magnitude of both theoretical and experimental
uncertainties.

In this situation it should be desirable to see if mean--field
models produce in light nuclei results similar to those obtained
for the medium--heavy ones. This would exclude 
explanations of the contradictory results such as the possibility
that the smallness of the global MEC  effect in medium--heavy nuclei
is due to the cancellations between the contributions of a large
amount of particle--hole excitations.

In order to investigate this point, we have applied to the $^4$He
nucleus the model we have used to study the quasi--elastic
responses in $^{12}$C and $^{40}$Ca \cite{ama94}.

In this model the ground state is described as a Slater
determinant of single particle wave functions produced by a
mean--field potential of Woods--Saxon type. The excited states
are built up as one particle--one hole (1p--1h) and 2p--2h
excitations, where the particle wave functions are obtained
solving the Schr\"odinger equation in the continuum with the
same Woods-Saxon potential.

Within this model we have evaluated the quasi--elastic response
functions as described in Ref. \cite{ama94}: the longitudinal
response is produced by the one--body charge operator, while the
transverse response is obtained adding to the one--body
convection and magnetisation currents, the two-body MEC.  These
have been calculated considering the so--called seagull or
contact, pionic or pion in flight, and $\Delta$--isobar terms.

In Table 1 we give the parameters of the Woods--Saxon potential
used in our calculations and defined as in Ref. \cite{ama94}. The
ground state properties of the $^4$He do not constrain the
spin--orbit part of the potential, which, on the other hand, 
affect the continuum single
particle wave functions used to calculate the responses. We have
studied the sensitivity of our results on the spin-orbit potential
using values taken from parametrizations considered in heavier
nuclei. We found that the effect on the responses is less than 1\%.
All the results presented in this report have been obtained 
using mean field potentials without
spin--orbit term. 

In Fig.1 we compare some results with the experimental data of
Ref. \cite{red90}.  The dashed lines have been obtained with a
Woods--Saxon potential, the WS1 of table 1, whose parameters
have been fixed in order to reproduce the energies of the 
$^4$He single particle levels. 
This is the usual procedure followed in
medium--heavy nuclei in order to choose the mean--field
parameters.  With this potential the charge distribution of
$^4$He is not very well reproduced, as it is shown in Fig. 2 by
the dashed line.

The parameters of the potential WS2 have been fixed to obtain
the best fit of the charge density, compatible with the
limitations of using a Woods-Saxon potential (the dashed--dotted
line of Fig.2). The results obtained with this potential are
presented in Fig.1 by the dashed--dotted lines.

The full lines of all the figures have been obtained with the
potential WS3 whose parameters have been fixed to obtain a
good agreement with the data of the longitudinal responses.
The values of single particle energies and of the charge
distributions obtained with this potential are rather different from
the experimental ones.

The longitudinal response functions are reasonably well
described by all the three calculations, while the transverse
response functions are always underestimated, in spite of the
fact that the MEC are included in the electromagnetic operator.

These results show a different trend with respect to the
medium--heavy nuclei where the ground state properties can
be described reasonably well with mean--field potentials. In the
present case, we could not reproduce simultaneously the various
ground state observables.
The $^4$He nucleus is too small to be reasonably
described by a mean--field model.

On the other hand, the aim of this work is not to produce a
realistic description of this nucleus, but rather to study the
possibility that MEC effects could be enhanced in few--body
systems.

Our main result is presented in Fig. 3, where the relative
differences between the transverse responses calculated with and
without MEC are shown for the three momentum transfer
considered. The left panels give the results corresponding to the
three different parametrizations of the Woods--Saxon potential
for $^4$He. The right panels show, for the same values of the
momentum transfer, the results obtained in $^{12}$C (full lines)
and in $^{40}$Ca (dashed lines) with the potentials WS1 of Refs.
\cite{ama93} and \cite{ama94}, respectively.

Three aspects shown in this figure deserve a comment.
\begin{enumerate}

\item In $^4$He, the contribution of the MEC at peak energies
is small, of the order of a few percent, if compared with the
full response. This result is rather independent from the
mean--field potential used.

\item The curves for $^4$He are very similar to those found for
$^{12}$C and $^{40}$Ca. In absolute value, at the peak energies, the
effect of the MEC becomes bigger the heavier is the nucleus.

\item The contribution of the MEC at the peak energies, with the
$\Delta$ isobar current included, is negative for transfer
momenta bigger than 400 MeV/c.

\end{enumerate}

These results show that the MEC contributions produced
by mean--fields models in $^4$He are similar to those
obtained in medium--heavy nuclei. The possibility of an
enhancement of these contributions in light nuclei
due to the smallness of these systems should be excluded.

It is worthwhile to point out the similarity of our $^4$He results
with the NT curves of Fig. 8 of Ref. \cite{lei90}.
Using a model quite different from ours, Leidemann and
Orlandini obtained these curves with purely central 
short--range correlations. They also show that the addition of the
tensor pieces of the correlation increases
the relative contribution of MEC up to 10--15\%, at the peak energy.

All these facts lead us to conclude that the small MEC effects
found in medium and heavy nuclei is due to the lack of short--range
correlations, and in particular their tensor components,
not taken into account in the mean--field models used to describe
these many--body systems.

One may claim that the contribution of MEC in heavy nuclei can be
enhanced by the presence of other effects which are usually not
considered, for example relativity and RPA long--range
correlations. 

We think relativistic effects are not playing an important role in
this context, because MEC contributions in light nuclei are large
even in non relativistic treatments. 
This idea is confirmed by a calculation done within the 
relativistic Fermi gas model by Blunden and Butler \cite{blu89} 
for the quasi elastic excitation of $^{40}$Ca where
MEC effects are evaluated to be of the order of a 10\%, but they are
not including the $\Delta$-isobar current. 
This is the same value we obtained with our model when we switch off
this component of the two body current
\cite{ama93,ama94}.

The role of RPA correlations on the MEC in the quasi--elastic region
is not clear. A recent work of the Gent group \cite{slu95} shows
considerable MEC effects, 20--30\% of the strength of the
quasi--elastic peak, within a non--relativistic
Hartree--Fock--RPA model. 

Full RPA calculations of MEC contributions performed at lower
energies, but at the same values of the momentum transfer, show
scarce sensitivity  to the RPA correlations \cite{ama92}.  
Furthermore, in continuum RPA calculations with
finite range residual interactions \cite{ama93,bub91} the
one--body quasi--elastic responses
do not show sizeable differences with mean--field results. 

In conclusion, we have shown that within
mean--field calculation the
MEC contribution in the quasi--elastic excitation of $^4$He 
is small, analogously to what happens in medium--heavy nuclei.
We deduce that this results is due to the lack, in the mean--field
approach, of short--range correlations. Calculations of MEC in
medium--heavy  nuclei with explicit treatment of
the short--range correlations are
desirable in  order to clarify definitively the problem.

This work has been partially supported by the agreement
between the C.I.C.Y.T. (Spain) and the I.N.F.N. (Italy).

\vskip 2 cm

{\bf Table.}
Parameters of the Woods--Saxon potential$^{\rm a}$
used in the various calculations described in the text and
single--particle energies obtained for the proton and neutron
$1s_{1/2}$ levels. The spin--orbit part is
switched off and the value of the Coulomb radius is taken equal
to the value of $R$. In the last row the experimental single
particle energies are shown.
\begin{center}
\begin{tabular}{cccccc}
      &   & $V$ & $R$ & $a$ & $\epsilon$ \\
      &   & $[MeV]$ & $[fm]$ & $[fm]$ & $[MeV]$ \\
\hline
 WS1   &  p & -65.83  &  1.70  & 0.60 & -19.52 \\
       &  n & -66.00  &  1.70  & 0.60 & -20.53 \\
\hline
 WS2   &  p & -52.11  &  1.80  & 0.20 & -17.24 \\
       &  n & -52.11  &  1.80  & 0.20 & -18.16 \\
\hline
 WS3   &  p & -55.00  &  1.98  & 0.85 & -17.39 \\
       &  n & -55.00  &  1.98  & 0.85 & -18.17 \\
\hline
 exp   &  p &         &        &      & -19.82 \\
       &  n &         &        &      & -20.58 \\
\hline
\end{tabular}
\end{center}
$^{\rm a}$See Ref. \cite{ama94} for the definition of
the potential.

\newpage
\noindent {\bf Figure Captions}
\vskip 1.5 cm

FIG.~1.~~
Longitudinal and transverse response functions
for different values of the momentum transfer. The dashed,
dashed--dotted and full lines have been calculated with the WS1,
WS2 and WS3 potential respectively. The experimental date have
been taken from Refs. \cite{red90}.

FIG.~2.~~
Charge densities obtained with the WS1 (dashed
line), WS2 (dashed--dotted line) and WS3 (full line) potentials
compared with the experimental one (Ref. \cite{mcc77}).

FIG.~3.~~
Relative differences between the transverse
responses calculated with and without MEC. The value
$\displaystyle \delta R_T =
\frac{R_T^{\rm OB+MEC} - R_T^{\rm OB}}{R_T^{\rm OB+MEC}}$ is 
plotted for different cases. In the left panels, we show the
results obtained for the three mean--field potentials considered
in this work for $^4$He. The curves are labelled as in Fig.1 and
2.  The right panels show the results for the WS1 potentials of Refs.
\cite{ama93} and \cite{ama94} for $^{12}$C (full lines)  and
$^{40}$Ca (dashed lines), respectively.


\begin{thebibliography}{99}

\bibitem{fro91}
A good review is provided by the section II of the book
{\it Modern Topics in Electron Scattering}, B. Frois and 
I. Sick eds., World Scientific, 1991.

\bibitem{lowen}
A. Arima, Y. Horikawa, H. Hyuga and T. Suzuki,
Phys. Lett. {\bf 40}, 1001 (1978);
J.F. Mathiot and B. Desplanques, 
Phys. Lett. {\bf B101}, 141 (1981);
T. Suzuki and H. Hyuga, Nucl. Phys. {\bf A402}, 491 (1983);
P.G. Blunden and B. Castel, Nucl. Phys. {\bf A445}, 742 (1985);
S. Krewald, A.M. Lallena and J.S. Dehesa, 
Nucl. Phys. {\bf A448}, 685 (1986).
 
\bibitem{ama92}
J.E. Amaro and A.M. Lallena, Nucl. Phys. {\bf A537}, 585 (1992).

\bibitem{mediu}
J.W. Van Orden and T.W. Donnelly,
Ann. Phys. (N.Y.) {\bf 131}, 451 (1981);
M. Kohno and N. Otsuka, Phys. Lett. {\bf B98}, 335 (1981);
W.M. Alberico, M. Ericson and A. Molinari,
Ann. Phys. {\bf 154}, 356 (1984);
W.M. Alberico, T.W. Donnelly and A. Molinari,
Nucl.Phys. {\bf A512}, 541 (1990);
M.J. Dekker, P.J. Brussaard and J.A Tjon, Phys. Lett. {\bf
B266}, 249 (1991); Phys. Rev. {\bf C49}, 2650 (1994).

\bibitem{blu89} P.G. Blunden and M.N. Butler, Phys. Lett. {\bf
B219}, 151 (1989)

\bibitem{ama93}
J.E. Amaro, G. Co' and A.M. Lallena, 
Ann. Phys. (N.Y.) {\bf 221}, 306 (1993).

\bibitem{ama94}
J.E. Amaro, G. Co' and A.M. Lallena, 
Nucl. Phys. A {\bf 578}, 365 (1994).

\bibitem{car91}
See e.g. J. Carlson, V.R. Pandharipande and R. Schiavilla,
in {\it Modern Topics in Electron Scattering}, B. Frois and 
I. Sick eds., World Scientific, 1991.

\bibitem{gro91}
See e.g. F. Gross,
in {\it Modern Topics in Electron Scattering}, B. Frois and 
I. Sick eds., World Scientific, 1991.

\bibitem{fri91}
See e.g. J.L. Friar,
in {\it Modern Topics in Electron Scattering}, B. Frois and 
I. Sick eds., World Scientific, 1991.

\bibitem{ros90}
S. Rosati, M. Viviani and A. Kievsky, 
Few-Body Syst. {\bf 9}, 1 (1990);
A. Kievsky, M. Viviani and S. Rosati, 
Nucl. Phys. {\bf A551}, 241 (1993).

\bibitem{car94}
J. Carlson and R. Schiavilla, Phys. Rev. Lett. {\bf 68}, 3682 (1992);
Phys. Rev. C {\bf 49}, R2880 (1994).


\bibitem{pie90}
S.C. Pieper, R.B. Wiringa and V.R. Pandharipande, 
Phys. Rev. Lett. {\bf 64}, 364 (1990);
G. Co', A. Fabrocini, S. Fantoni and I.E. Lagaris, 
Nucl. Phys. A {\bf 549}, 439 (1992);
G. Co', A. Fabrocini and S. Fantoni  Nucl. Phys. A {\bf 568}, 73 (1994).

\bibitem{red90}
K.F. von Reden et al. , Phys. Rev. C {\bf 41}, 1084 (1990);
A. Zghiche et al. , Nucl. Phys. A {\bf 572}, 513 (1994).

\bibitem{mcc77} 
J.S. McCarthy, I. Sick and R.R. Whitney, Phys. Rev. C {\bf 15}, 1396 (1977).

\bibitem{lei90}
W. Leidemann and G. Orlandini, Nucl. Phys. {\bf A506}, 447 (1990).


\bibitem{slu95}
V. Van der Sluys, J. Ryckebusch and M. Waroquier, Phys. Rev. C
{\bf 51}, 2664 (1995).

\bibitem{bub91}
M. Buballa, S.Dro\.zd\.z, S.Krewald and J.Speth, Ann. Phys. {\bf
208}, 346 (1991).

\end{thebibliography}
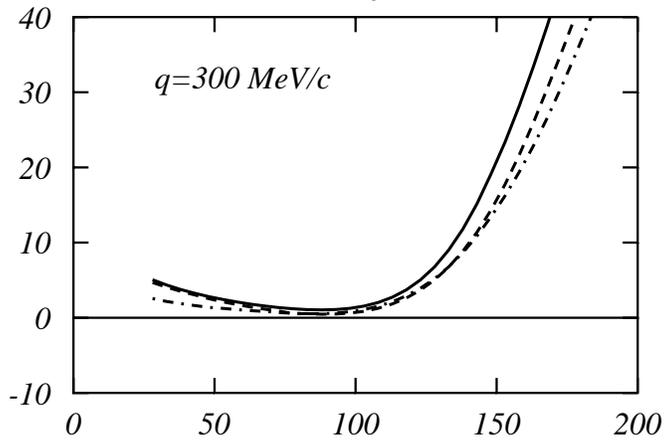
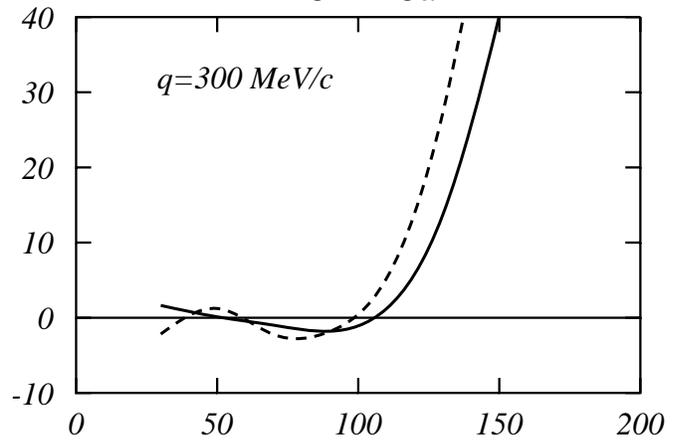
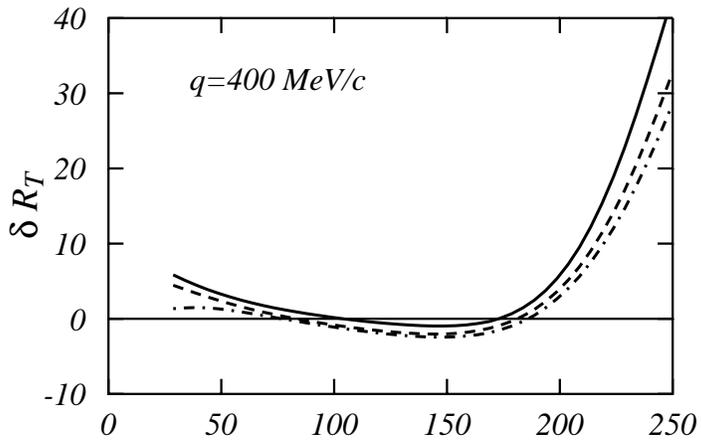
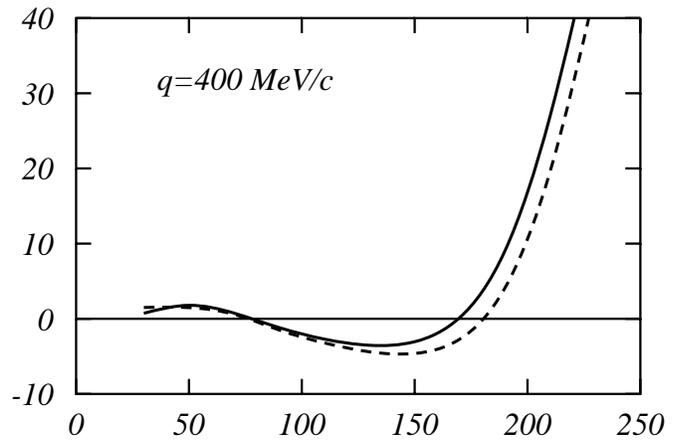
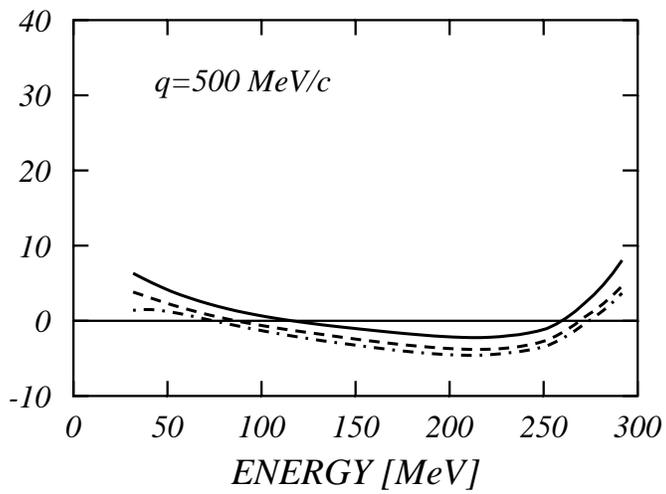
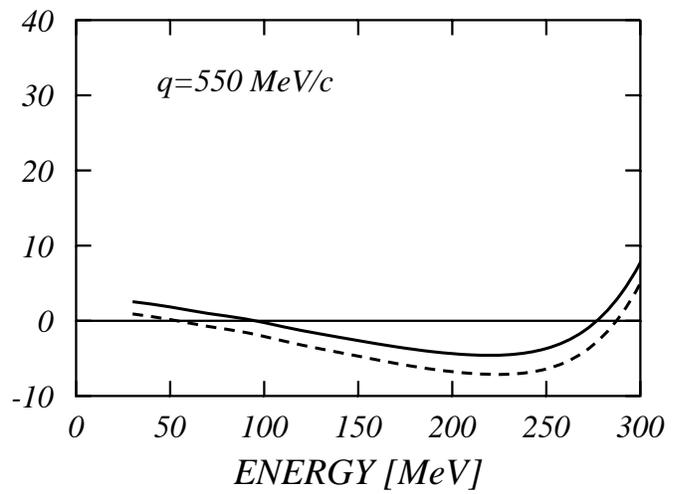
\end{document}